\documentclass[times,twocolumn,final]{elsarticle}

\usepackage{medima}
\usepackage{framed,multirow}

\usepackage{amsmath,amsfonts}
\usepackage{array}
\usepackage{textcomp}
\usepackage{stfloats}
\usepackage{url}
\usepackage{verbatim}
\usepackage{graphicx}
\usepackage{bm}
\usepackage{amsmath}
\usepackage{amssymb}
\usepackage{comment}
\usepackage{algorithm}
\usepackage{algpseudocode}
\usepackage{graphicx}
\usepackage{color}
\usepackage{url}
\usepackage{multirow}
\usepackage{booktabs}
\usepackage{multirow}
\usepackage{wrapfig}
\usepackage{enumerate}
\usepackage{caption}
\usepackage{cite}
\usepackage{bbding}
\usepackage{textcomp}
\usepackage{pifont}
\usepackage{makecell}

\usepackage{amssymb}
\usepackage{latexsym}
\usepackage{hyperref}

\usepackage{url}
\usepackage{xcolor}
\hypersetup{hidelinks}

\definecolor{newcolor}{rgb}{.8,.349,.1}

\usepackage{mathtools}
\usepackage{amsmath}

\usepackage{hyperref}
\usepackage{booktabs}
\usepackage{multirow}
\journal{Medical Image Analysis}
\makeatletter\def\stmtitleheader{}\gdef\@journal{Preprint}\makeatother

\begin{document}

\verso{J.
Hao \textit{et~al.}}

\begin{frontmatter}

\title{Temporal Diffusion Bridge Model for Vessel Extraction Using Physics-informed Synthetic Data}

\author[1]{Jinkui \snm{Hao}}
\author[1]{Donald R. \snm{Cantrell}}
\author[1]{Ramez \snm{Abdalla}}
\author[1]{Sameer A. \snm{Ansari}}
\author[1]{Bo \snm{Zhou}\corref{cor1}}
\cortext[cor1]{Bo Zhou: \textit{bo.zhou@northwestern.edu}}
\address[1]{Department of Radiology, Northwestern University, Chicago, IL, USA}

\begin{abstract}
Accurate artery extraction from X-ray Coronary Angiography (XCA) is essential for the diagnosis and interventional treatment of coronary artery disease. However, reliable vessel extraction remains challenging due to low vessel visibility and motion artifacts caused by respiration and cardiac activity, under which conventional digital subtraction angiography methods often fail. Deep learning-based approaches typically rely on large, manually annotated datasets to achieve robustness in complex clinical settings. Moreover, most existing methods produce only binary vessel masks and neglect intra-arterial contrast information that is critical for clinical evaluation.
To address these limitations, we formulate vessel extraction as an image translation problem and propose a Temporal Diffusion Bridge Model (TDBM) trained on physics-informed synthetic data {without requiring manual vessel annotations.} The model predicts contrast-free XCA frames from contrast-enhanced sequences, enabling high-fidelity vessel recovery. Training data are generated using a physics-based XCA synthesis engine that produces large-scale, diverse, and perfectly aligned pairs of contrast-filled and contrast-free images. Leveraging this data, the proposed temporally conditioned diffusion bridge learns cross-domain mappings while incorporating adjacent multi-frame context to improve temporal consistency. In addition, a vessel-aware contrastive learning loss is introduced using background-invariant image pairs from the synthesis engine to enhance fine-grained vascular discrimination.
The proposed framework is evaluated on two XCA vessel segmentation datasets and further validated through a reader study on two real patient XCA sequence datasets. Furthermore, zero-shot vessel segmentation experiments are conducted on a retinal imaging dataset and a fluorescence microscopy dataset. Experimental results demonstrate that the proposed method consistently outperforms existing approaches, while exhibiting strong generalization capability in zero-shot segmentation scenarios.
Code and trained model weights will be available at:~\href{https://github.com/Advanced-AI-in-Medicine-and-Physics-Lab/TDBM}{https://github.com/Advanced-AI-in-Medicine-and-Physics-Lab/TDBM}.
\end{abstract}


\end{frontmatter}

\section{Introduction}
Coronary artery disease (CAD), one of the most prevalent cardiovascular disorders, poses a major threat to human health worldwide~\citep{martin20252025}. Accurate diagnosis of CAD relies heavily on precise characterization of coronary arteries~\citep{mahabadi2010association}.
X-ray coronary angiography (XCA) is a primary imaging modality for assessing the morphological and functional characteristics of coronary arteries, playing a key role in both diagnosis and interventional treatment of CAD. However, vessel visibility in XCA is often degraded by the superimposed fluoroscopic densities of bones and soft tissues. Reliable vessel extraction from XCA is therefore essential for downstream automated analyses, such as interventional path planning and quantitative stenosis assessment.
Conventional Digital Subtraction Angiography (DSA) techniques, which aim to isolate vessel structures, fail in XCA due to respiratory and cardiac motion artifacts~\citep{jin2017extracting}.

Developing a reliable and robust automated method for vessel extraction from XCA has thus attracted considerable attention. Existing approaches can be broadly categorized into two groups. The first comprises training-free methods, such as edge- or contour-based tracking~\citep{brieva2004coronary, lalinia2024coronary} and principal component analysis (PCA)-based algorithms~\citep{zhang2018vesselness, qin2022robust,fu2024robust,wu2025denver}. Although these methods do not require annotated data, they exhibit limited generalizability because numerous parameters must be manually tuned for varying imaging conditions and anatomical differences.
The second group includes deep learning–based methods, which train data-driven models using manually annotated vessel masks as ground truth~\citep{yang2019deep, thuy2021coronary, lee2024real,he2025deep}. Despite their success, these approaches typically require large-scale labeled datasets to achieve sufficient robustness across diverse acquisition settings, which poses a significant practical challenge.
To alleviate the dependence on labeled data, several recent studies have explored synthetic data generation for model training~\citep{he2025conditional,ge2026generative}. However, existing approaches often suffer from one or more limitations:
1) they require annotated data to train the generative model;
2) the domain gap between synthetic and real data leads to inferior generalization; or
3) few-shot fine-tuning with real labels is still needed to bridge domain gap.
Moreover, most deep learning-based methods focus solely on binary vessel segmentation and overlook intra-arterial contrast information, which is essential for clinical interpretation and hemodynamic analysis.

In this work, we present a synthetic data-driven temporal diffusion bridge framework for vessel extraction in XCA. Instead of performing direct vessel segmentation, we formulate the task as an image-to-image translation problem, predicting contrast-free images from contrast-filled sequences, to preserve intra-arterial contrast dynamics.
Specifically, we first introduce a physics-based XCA synthesis engine capable of generating diverse and realistic pairs of contrast-filled and contrast-free images. 
Unlike generative models that rely on data-driven priors, our physics-based synthesis reproduces the X-ray formation process through 3D vessel modeling, contrast injection simulation, and digitally reconstructed radiographs (DRRs), resulting in anatomically diverse and physically realistic XCA-like images.
Building on this, we design a Temporal Diffusion Bridge Model (TDBM) trained with 50,000 synthetic samples to learn the mapping between contrast-filled and contrast-free domains while leveraging multi-frame temporal context for enhanced sequence consistency.
Furthermore, we employ the synthesis engine to generate background-invariant vessel pairs and propose a vessel-aware contrastive loss to strengthen fine-grained vascular discrimination.
We perform a comprehensive evaluation combining quantitative benchmarks with qualitative clinical assessments. Specifically, vessel segmentation performance is validated on two benchmark datasets, while a reader study on both internal and external real XCA sequences datasets further assesses noise suppression, vessel fidelity, temporal consistency, and clinical utility.
Experimental results demonstrate that our approach consistently outperforms state-of-the-art methods, including fully supervised models trained on real data.
Beyond XCA, we demonstrate the distinct advantage of our physics-driven approach by conducting zero-shot vessel segmentation on retinal and fluorescence microscopy images. In these experiments, the model exhibits strong generalization capabilities, accurately extracting vessels from unseen modalities.

\section{Related Work}
This section contextualizes our proposed approach by reviewing the evolution of vessel extraction techniques in XCA and examining recent paradigms in synthetic data generation for medical imaging.

\subsection{Evolution of Vessel Extraction in XCA}
The extraction of coronary arteries from XCA sequences has evolved from geometry-based filtering to data-driven deep learning.

Early methodologies relied on mathematical formulations to distinguish vascular structures from complex backgrounds. Hessian-based strategies, particularly the Frangi filter~\citep{frangi1998multiscale}, became the standard for enhancing tubular structures. Recent adaptations of this principle, such as the Enhanced Frangi Filter (EFFCA)~\citep{wang2023effca} and anisotropic diffusion hybrids~\citep{bhattacharya2013optimized}, have focused on optimizing edge preservation and contrast for clinical deployment.
Parallel to filtering, background subtraction techniques emerged to handle the dynamic noise inherent in XCA. Sparse decomposition methods, including Vesselness-constrained Robust PCA (RPCA)~\citep{zhang2018vesselness} and Multi-Component RPCA~\citep{qin2022extracting}, attempt to mathematically decouple the foreground vessels from skeletal and catheter interference. However, these unsupervised methods often lack the generalization capability required for challenging cases with severe vessel overlap or motion artifacts.

The advent of Convolutional Neural Networks (CNNs) marked a shift toward supervised segmentation. Initial works, such as the U-Net adaptations by Yang et al.~\citep{yang2019deep} and Kaba et al.~\citep{kaba2023application}, demonstrated superior performance over classical filters but largely treated XCA videos as independent 2D frames. This frame-by-frame approach neglects critical temporal correlations, rendering segmentation susceptible to motion blur and inconsistent vessel delineation.
To address this, recent architectures have integrated spatiotemporal modules. Solutions like STCA-Net~\citep{gao2025spatio}, TVS-Net~\citep{he2025deep}, and TCA-Net~\citep{yang2025temporal} leverage adjacent frames to enforce temporal smoothness. Similarly, Tang et al.~\citep{tang2025temporally} demonstrated that incorporating temporal cues significantly improves the stability of main coronary artery segmentation.

{GAN-based image-to-image translation methods, such as pix2pix~\citep{isola2017image}, learn a direct mapping between source and target domains, but their adversarial training may suffer from instability and can be less suitable when subtle image structures must be preserved. Diffusion models~\citep{ho2020denoising} provide a more stable generative framework, but conventional diffusion typically learns a noise-to-data process rather than a direct transition between two image domains. Diffusion bridge models address this limitation by constructing a stochastic path directly between paired source and target distributions. In particular, Brownian Bridge Diffusion Models (BBDM)~\citep{li2023bbdm} define a finite bridge between two image domains, making them well suited to image-to-image translation with relatively small structural differences. Related approaches such as $I^2$SB~\citep{liu2023i2sb} formulate image translation through a Schr\"odinger bridge with entropy-regularized optimal transport. In this work, we adopt the Brownian-bridge formulation because the contrast-enhanced and contrast-free XCA frames share most anatomical background structures, allowing the bridge to primarily model the transient vessel-related contrast while preserving the underlying anatomy. We further extend this formulation with temporal conditioning and vessel-aware representation learning for XCA vessel extraction.}

{Recent foundation-model-based segmentation methods, including SAM and its medical adaptations~\citep{kirillov2023segment,ma2024segment}, have also shown strong generalization across segmentation tasks. Several recent studies have adapted such models to vascular and coronary angiography segmentation, including CoroSAM~\citep{ferrari2025corosam} and related approaches. However, these methods generally rely on annotated images for task-specific adaptation or prompting, whereas our method focuses on annotation-free vessel extraction without requiring pixel-level vessel labels.}

Despite these architectural advances, the reliance on massive, pixel-wise annotated datasets remains a primary barrier. While semi-supervised frameworks utilizing focal masking~\citep{wang2024semi} or teacher-student networks~\citep{lee2024real} have attempted to mitigate this by leveraging unlabeled data, they still depend on a core set of high-quality real-world annotations, which are notoriously difficult to acquire for XCA.

\begin{figure*}[!h]
    \centering{  
    \includegraphics[width=1\linewidth]{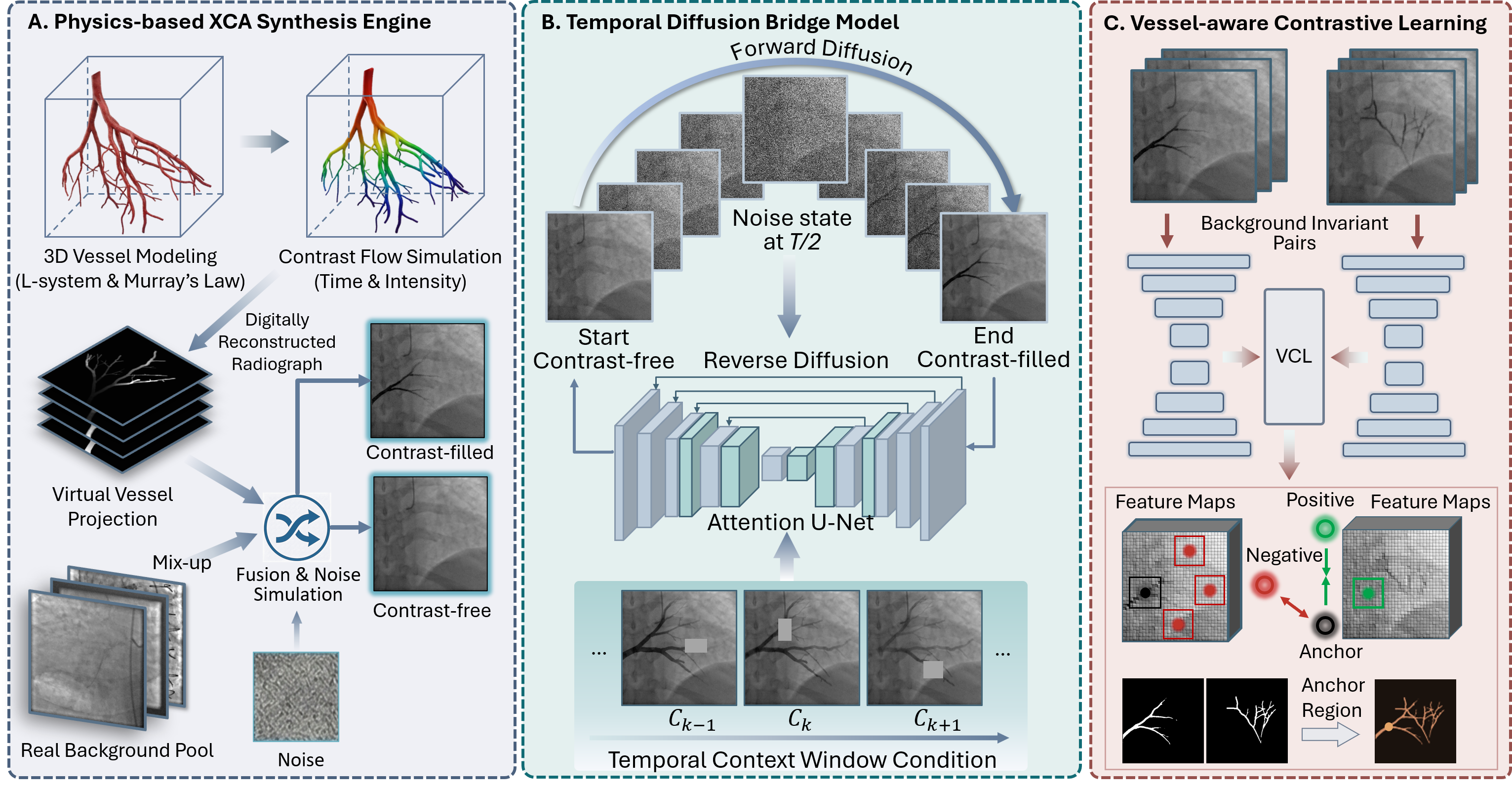}
    }
    \caption{Framework overview. (A) A physics-based engine synthesizes realistic, aligned training pairs. (B) The Temporal Diffusion Bridge Model (TDBM) performs vessel extraction as a Brownian Bridge process, using an Attention U-Net conditioned on a temporal window for motion consistency. (C) Training is enhanced by a vessel-aware contrastive loss (VCL) for feature discrimination.}
    \label{fig-framework}
\end{figure*}

\subsection{Medical Image Synthesis and Generalization}
To overcome the scarcity of annotated medical data, research has increasingly pivoted toward training models on synthetic data.

\textbf{Generative and Simulation Frameworks:} Early adoption of synthetic data utilized Generative Adversarial Networks (GANs) to augment real datasets, as seen in brain MRI tumor segmentation~\citep{shin2018medical}. More recently, the focus has shifted toward physics-based simulation and domain randomization to create "free" labels. The SyntheX framework~\citep{gao2023synthetic} and work by Kreitner et al.~\citep{kreitner2024synthetic} demonstrated that models trained on simulated X-ray and OCT data could rival those trained on real clinical data. This trend toward generalization is further exemplified by AnyStar~\citep{dey2024anystar} and broadly applicable biomedical volume representations~\citep{dey2024learning}, which achieve zero-shot or few-shot capabilities via large-scale synthetic pretraining.

\textbf{Diffusion and Vascular Synthesis:} Diffusion probabilistic models have recently emerged as a powerful tool for generating high-fidelity medical images. For instance, noise-consistent diffusion frameworks have shown promise in synthesizing realistic polyp images to boost segmentation performance~\citep{qiu2025noise}.
However, the application of these advanced synthesis techniques to XCA remains underexplored. While Duan et al.~\citep{duan2024training} proposed a pipeline for DSA, and Hao et al.~\citep{hao2026dosta} explored domain-shuffling for XCA, these methods often fail to capture the complex, physics-based temporal dynamics of coronary flow. Furthermore, generative approaches like those by He et al.~\citep{he2025conditional} show promise in few-shot scenarios but often lack the precise anatomical control required for robust clinical validation.

\section{Proposed Method} 
\label{sec:method}
In this section, we detail our proposed method, including a physics-based data engine for XCA synthesis, the TDBM, and the vessel-aware contrastive learning for the model training.

\subsection{Physics-based Data engine for XCA Synthesis}
As shown in Fig.~\ref{fig-framework}a, to overcome the limitations of purely data-driven synthesis models (e.g., GANs or diffusion models), we developed a Physics-informed Data Engine. This engine is designed to ensure the anatomical and physiological realism of synthesized XCA images by mechanistically modeling the entire angiographic formation process, from biophysically constrained vascular morphology to X-ray attenuation physics. Our approach comprises four key modules: 3D vessel synthesis, contrast flow simulation, physics-based 3D-to-2D projection, and realistic XCA image fusion.

We first construct realistic vascular geometries using a rule-driven Lindenmayer system~\citep{zamir2001arterial} that procedurally encodes the hierarchical branching structure. 
Critically, the process is physiologically constrained: stochastic parameters are sampled within ranges dictated by human anatomy, and bifurcation angles and diameter tapering are rigorously computed based on Murray’s law~\citep{murray1926physiological}-derived relationships. This ensures mass conservation and realistic hydrodynamic resistance across all vessel levels, guaranteeing an anatomically plausible and hemodynamically coherent vessel tree embedded in the 3D space.

To accurately emulate the dynamic propagation of the contrast agent, we model the intensity of each vessel voxel as a time-dependent concentration profile constrained by physiological flow and contrast dynamics.
We establish a segment-dependent flow velocity $v_s$ that is hemodynamically coherent with the vessel diameters dictated by Murray's Law. This ensures realistic flow physics across the vascular hierarchy. The arrival time $t_0(r)$ for contrast to reach position $r$ is then calculated via the cumulative path integral of these segment velocities from the injection site:
$$
t_0(r) = \sum_{s \in \text{path}(r)} \frac{L_s}{v_s}
$$
where $L_s$ is the length of vessel segment $s$. This approach accurately captures the differential propagation speeds inherent to the vascular anatomy.
To accurately capture the non-symmetric temporal characteristics of contrast enhancement, including the critical physiological washout tail, we adopt the Gamma-variate function~\citep{benner1997accuracy} to model the local intensity $I(r, t)$:
$$
I(r, t) = A \cdot (t - t_0(r))^{\alpha} e^{-\frac{t - t_0(r)}{\beta}}, \quad \text{for } t \ge t_0(r)
$$
Here, $A$ scales the maximal enhancement, while $\alpha$ and $\beta$ govern the temporal dispersion and clearance rate. The resulting 4D vascular field provides a rigorous description of contrast density evolution.

The simulated vascular field is discretized into volumetric voxel grids through a voxel traversal algorithm derived from Cohen’s line-voxel intersection method~\citep{cohen20023d}. Each vessel segment is rasterized according to its diameter and intensity, producing anatomically continuous 3D angiographic volumes. To mimic X-ray image formation, these 3D volumes are projected onto a virtual detector using DRR~\citep{gopalakrishnan2022fast}. For each frame, the DRR integrates attenuation values along projection rays:
\begin{equation}
I_{\text{DRR}}(x,y) = \int_{\text{ray}(x,y)} I(r,t),dr,
\end{equation}
where $(x,y)$ denotes the detector coordinates. Randomized camera rotations and translations are applied to simulate variation in fluoroscopic viewpoints and patient positioning.

The projected vascular images are fused with contrast-free fluoroscopic backgrounds. A pool of real XCA frames without visible vessels serves as the background source. Two randomly sampled backgrounds are linearly combined via MixUp~\citep{zhang2018mixup} to generate diverse texture patterns, and random elastic deformations and gamma transformations are optionally applied as data augmentation to increase variability in background geometry and exposure. The vessel-only projection $V_t$ and the composite background $B$ are fused as:
\begin{equation}
I_t = B - \alpha V_t,
\end{equation}
where $\alpha$ controls vessel visibility and contrast balance. The fused image is further perturbed with randomly sampled Poisson noise to emulate the noise characteristics of real image acquisition.
This physically grounded synthesis process produces realistic, temporally consistent images that preserve vessel morphology, flow dynamics, and imaging artifacts, thereby providing high-fidelity data for downstream model training.

\subsection{Temporal Diffusion Bridge Model}

{As shown in Fig.~\ref{fig-framework}b, in the TDBM, we formulate vessel extraction as a conditional generation task. We construct a diffusion bridge between a background-only image $x_0 \sim q_{\text{data}}(I_B)$ and its corresponding vessel-containing XCA image $y \sim q_{\text{data}}(I_A)$, where $x_0,y \in \mathbb{R}^{1\times256\times256}$.}

\textit{Forward Process.}
We define a discrete-time Brownian Bridge process $\{x_t\}_{t=0}^T$~\citep{li2023bbdm}. Given the start point $x_0$ and end point $y$, the distribution of $x_t$ at any intermediate timestep $t \in [0, T]$ is a Gaussian distribution that can be sampled analytically:
\begin{equation}
q(x_t|x_0, y) = \mathcal{N}(x_t; \mu_t, \sigma_t^2 I)
\label{eq:bb_mean_var}
\end{equation}
where $\mu_t = (1-m_t)x_0 + m_t y$ and $\sigma_t^2 = 2s\,(m_t - m_t^2)$.
Following BBDM, $m_t$ increases linearly from $0.001$ to $0.999$ over $T=1000$
steps and $s=1$, so that $x_t \to x_0$ at $t=0$, $x_t \to y$ at $t=T$, and the
variance is maximal at the midpoint. A sample is obtained as
$x_t = (1-m_t)x_0 + m_t y + \sigma_t\varepsilon$, $\varepsilon\sim\mathcal{N}(0,I)$.

\textit{Reverse Objective and Conditioning.} To capture the dynamics of XCA
sequences, the denoiser is conditioned on a temporal context window
$C_k = \{I_A^{(k-\Delta)},\dots,I_A^{(k+\Delta)}\} \in \mathbb{R}^{(2\Delta+1)\times 256\times 256}$,
which contains the target contrast-filled frame $y^{(k)} = I_A^{(k)}$ and its
neighbours. Importantly, the contrast-free frame $x_0^{(k)} = I_B^{(k)}$ is used
only to construct $x_t$ and the regression target during training and is never
provided as network input, since it is unavailable at inference. 
The network $f_\theta$, an attention U-Net, takes the channel-wise concatenation
$[x_t, C_k] \in \mathbb{R}^{(2\Delta+2)\times 256\times 256}$ and the timestep $t$,
and predicts the bridge residual $x_t - x_0 = m_t(y - x_0) + \sigma_t\varepsilon$,
following the BBDM objective. The model is optimized with an MSE objective:
\begin{equation}
\mathcal{L}_{\text{bridge}}(x_0, y, C_k) =
\mathbb{E}_{t,\varepsilon}\Big\| m_t\big(y - x_0\big) + \sigma_t\varepsilon
- f_\theta\big([x_t, C_k], t\big) \Big\|_2^2 .
\end{equation}

\subsection{Training with Vessel-aware Contrastive Learning}

Building on the TDBM, which learns the mapping between contrast-filled and contrast-free domains, we further introduce a vessel-aware contrastive learning scheme to enhance the model’s sensitivity to vascular structures, as shown in Fig.~\ref{fig-framework}c. The physics-based synthesis engine enables the generation of paired XCA images that share identical backgrounds but contain distinct vascular patterns. These pairs provide a controlled setting for learning vessel-relevant representations while being invariant to background variations.

Given a paired input $(I_A^{(1)}, I_A^{(2)})$, both images are passed through the TDBM, and feature maps are extracted from an intermediate convolutional block at level $l$. Since the vessel masks of both images are known from synthesis, we define the vessel region $\Omega_v$ as the union of vessel areas from both images, and the complement $\Omega_b$ as the background region. 
{By setting $\Omega_v=\Omega_v^{(1)}\cup\Omega_v^{(2)}$ and $\Omega_b=\Omega_v^c$, all negatives are guaranteed to be vessel-free in both views. Using the intersection instead would incorrectly treat vessel pixels present in only one view as background negatives, which would corrupt the InfoNCE objective.
The two vessel trees are independently generated, so their intersection is also typically small and would provide very few meaningful anchors. With union-based anchors, the loss encourages features at vessel-bearing locations to remain close to the corresponding location in the paired view while being separated from genuinely vessel-free background, thereby promoting invariance to the specific vessel realization.}
From the feature map of image 1, features within $\Omega_v$ are treated as \textit{anchors}, their spatially corresponding features in image 2 as \textit{positives}, and features sampled from $\Omega_b$ as \textit{negatives}. 
The vessel-aware contrastive loss is formulated as:
\begin{equation}
\mathcal{L}_{\text{vcl}}
=\\ 
- \log
\frac{\exp(\text{sim}(f_a, f_p)/\tau)}
{\exp(\text{sim}(f_a, f_p)/\tau)
+ \sum_{n\in \Omega_b}\exp(\text{sim}(f_a, f_n)/\tau)},
\end{equation}
where $f_a$, $f_p$, and $f_n$ denote the anchor, positive, and negative feature vectors, $\text{sim}(\cdot)$ is the cosine similarity, and $\tau$ is a temperature parameter.
By explicitly contrasting vessel and background features under shared imaging conditions, the loss makes the network amplify responses to vessel patterns in the intermediate feature space and treat them as components to be suppressed; consequently, when inferring back to the background domain, vessel residues are removed more effectively, artifacts are reduced, and background consistency around former vessel regions is improved.
{VCL is applied at five levels spanning both the encoder and decoder paths: the last residual block of the deepest encoder stage and the bottleneck output (both 512 channels at $1/8$ resolution), followed by the three decoder stages at $1/4$, $1/2$, and full resolution (512, 256, and 128 channels, respectively). These levels were selected to capture complementary information: deeper features encode larger vessel segments and their context, whereas shallower decoder features better preserve thin distal branches.}

Although the combination of synthetic data and the TDBM with vessel-aware contrastive learning yields promising performance, we observe that a residual domain gap between synthetic and real XCA data may still hinder robustness on some clinical cases. To mitigate this, a small portion of real XCA sequences is incorporated into training. For each real sequence, the first frame is treated as the contrast-free reference (i.e., the ground truth), while the subsequent frames serve as contrast-filled inputs. However, due to cardiac motion and acquisition variations, the real data are not perfectly aligned, introducing noise into supervision.
To balance this trade-off, we assign an annealing loss to the real-data, which has a small initial weight $\lambda(t)$ and gradually decrease it to zero during training. This annealing strategy allows the model to benefit from domain adaptation in early stages while avoiding the propagation of misalignment noise later. The overall training objective combines the synthetic supervision, contrastive, and real-data supervision terms as:
\begin{equation}
\mathcal{L}_{\text{total}} 
= \mathcal{L}_{\text{sup}}^{\text{syn}} 
+ \alpha \mathcal{L}_{\text{vcl}}
+ \lambda(t)\, \mathcal{L}_{\text{sup}}^{\text{real}},
\quad \text{where}\;\lambda(t)\!\downarrow\!0.
\end{equation}
where $\alpha$ represents the weighting factors for the loss terms, and both $\mathcal{L}_{\text{sup}}^{\text{syn}}$ and $\mathcal{L}_{\text{sup}}^{\text{real}}$ correspond to MSE losses.

\subsection{Inference and vessel extraction}
\label{sec:inference}
{
Algorithm~\ref{alg:inference} summarizes the inference procedure of TDBM. For each contrast-enhanced frame $I_A^{(k)}$, a 3-frame temporal window ($\Delta=1$) is constructed and the center frame is used as the terminal bridge state, $x_T=I_A^{(k)}$. This is consistent with the training formulation because the terminal distribution of the Brownian bridge converges to the contrast-enhanced endpoint.
The learned reverse process is then applied for $K$ sampling steps to predict the corresponding contrast-free frame $\hat{I}_B^{(k)}$. The vessel-density map is obtained by subtraction,
\begin{equation}
\hat{V}^{(k)}
=
\mathrm{clip}\left(
\hat{I}_B^{(k)}-I_A^{(k)}
\right).
\end{equation}
This continuous map preserves intra-arterial contrast information and is used directly in the reader study. For quantitative segmentation evaluation, it is binarized using a fixed global threshold,
\begin{equation}
\hat{M}^{(k)}
=
\mathbb{I}
\left[
\hat{V}^{(k)}>\theta_{\mathrm{bin}}
\right],
\end{equation}
where $\theta_{\mathrm{bin}}=20/255$ is used for all datasets. No dataset-specific threshold tuning or additional morphological post-processing is applied.
}

\begin{algorithm}[!h]
\caption{{Inference of TDBM for vessel extraction}}
\label{alg:inference}
\begin{algorithmic}[1]
\Require Contrast-enhanced sequence $\{I_A^{(k)}\}_{k=1}^{N}$, trained model $f_\theta$,
         sampling timesteps $T = t_K > t_{K-1} > \cdots > t_1 = 0$
\Ensure Vessel-density maps $\{\hat{V}^{(k)}\}$ and binary masks $\{\hat{M}^{(k)}\}$
\State Resize all frames to $256\times256$ with intensities in $[0,1]$
\For{$k = 1,\ldots,N$}
    \State $C_k \gets \{I_A^{(k-1)}, I_A^{(k)}, I_A^{(k+1)}\}$
           \Comment{boundary frames repeated at sequence ends}
    \State $x_{t_K} \gets I_A^{(k)}$
    \For{$j = K,\ldots,1$}
        \State $\hat{\xi}_{t_j} \gets f_\theta\big(\mathrm{concat}(x_{t_j}, C_k),\, t_j\big)$
        \State $\hat{x}_0 \gets x_{t_j} - \hat{\xi}_{t_j}$
        \If{$j > 1$}
            \State Sample $x_{t_{j-1}} \sim p\big(x_{t_{j-1}} \mid x_{t_j}, \hat{x}_0, I_A^{(k)}\big)$
                   \Comment{reverse bridge transition}
        \EndIf
    \EndFor
    \State $\hat{I}_B^{(k)} \gets \hat{x}_0$
    \State $\hat{V}^{(k)} \gets \mathrm{clip}\big(2(\hat{I}_B^{(k)} - I_A^{(k)}),\,0,\,1\big)$
    \State $\hat{M}^{(k)} \gets \mathbb{I}\big[\hat{V}^{(k)} > \theta_{\mathrm{bin}}\big]$,
           \quad $\theta_{\mathrm{bin}} = 20/255$
\EndFor
\State \Return $\{\hat{V}^{(k)}\}, \{\hat{M}^{(k)}\}$
\end{algorithmic}
\end{algorithm}

\section{Study Data and Evaluation}
This evaluation in this study encompasses three major components: (1) vessel segmentation from XCA images for quantitative evaluation, (2) a reader study to assess clinical usability and interpretability, and (3) zero-shot segmentation experiments to evaluate cross-domain generalization on unseen imaging modalities.

\subsection{Vessel Segmentation}
To assess the quantitative vessel segmentation performance, we conducted experiments on both internal and external datasets with ground-truth annotations.

\textbf{Internal Dataset:}
The internal evaluation utilized a synthetic dataset of 30 sequences (7 frames/sequence), which was specifically designed to analyze the model's capability in leveraging temporal information. All images were resized to 256 × 256 pixels for consistency and computational efficiency.  The quantitative segmentation evaluation on the internal dataset is  performed at the peak-contrast frame.

\textbf{External Dataset:}
For external validation, we employed the public XCA30 dataset~\citep{hao2020sequential}, which contains 30 real XCA sequences (4 frames/sequence) with ground-truth annotations provided for the third frame. This benchmark allowed for objective comparison with state-of-the-art methods under standardized conditions.
{We additionally used DCA1~\citep{cervantes2019automatic}, comprising 134 single-frame XCA images with manual vessel annotations, as a second external benchmark to enable comparison with recent methods reported on this widely used dataset.}

Four standard metrics were calculated for both datasets: Accuracy, Sensitivity, Specificity, and the Dice Similarity Coefficient (DSC). These metrics provide comprehensive quantitative assessment of vessel segmentation performance.

\subsection{Reader Study}
Beyond quantitative metrics, we recognized that standard vessel segmentation measures do not fully capture the diagnostic interpretability, image fidelity, and temporal coherence prioritized by clinicians. Therefore, a comprehensive reader study was conducted to assess the method's practical utility and readiness for clinical scenarios.

\textbf{Internal Dataset:}
The reader study utilized an in-house dataset collected at Northwestern University, comprising 25 real XCA sequences (1,542 total frames; $\approx$ 62 frames/sequence) acquired at 15 fps. 

\textbf{External Dataset:}
To further validate clinical utility across different data sources, we use the XCA sequences from the public XACV dataset~\citep{wu2025denver}. The XACV dataset comprises 111 complete coronary artery X-ray videos from 59 patients, with each video containing an average of 86 frames. We randomly selected 15 sequences from XACV, which served as an external benchmark for the reader study.
{It is noted that all the Internal external test sets were not used for model tuning.}

\subsection{Zero-Shot Segmentation}
To validate the generalization capability and domain adaptability of our proposed method, we performed zero-shot vessel segmentation experiments on two distinct, out-of-distribution domains without any fine-tuning or domain-specific training.

\textbf{DRIVE}~\citep{staal2004ridge} is a widely-used benchmark for retinal vessel segmentation. 
The dataset was split into 20 images for training supervised baseline methods and 20 images for testing. Our zero-shot approach utilized only the 20 test images without any prior training on retinal images.

\textbf{VessMap}~\citep{viana2025new} is a fluorescence microscopy images dataset for measuring blood vessel segmentation performance under distribution shifts.
60 images were allocated for training supervised baselines, while our method performed zero-shot inference on the remaining 40 test images. 

\section{Experiments}

All experiments were conducted on a Linux server equipped with Tesla A100 GPUs. 
The model was trained using a distinct, large-scale dataset composed of 50,000 synthetically generated sequences with paired annotations, combined with 200 real XCA sequences that were incorporated using annealing loss learning strategies.
Table~\ref{tab:hyperparameters} summarizes the training and inference hyperparameters of TDBM.

\begin{table}[h]
\centering
\caption{{Key hyperparameters used for TDBM training and inference.}}
\begin{tabular}{ll}
\hline
Parameter & Value \\
\hline
Input resolution & $256\times256$ \\
Temporal radius $\Delta$ & $1$ (3 frames) \\
Diffusion timesteps $T$ & $1000$ \\
Reconstruction loss & MSE \\
VCL weight $\alpha$ & $0.01$ \\
VCL temperature $\tau$ & $0.05$ \\
VCL anchors / negatives & $256$ / $100$ \\
Optimizer & Adam \\
Initial learning rate & $5\times10^{-5}$ \\
Batch size  & $4$ \\
Training schedule & $30\times2500$ iterations \\
\hline
\end{tabular}
\label{tab:hyperparameters}
\end{table}

\begin{table*}[ht]
\scriptsize
\centering
\setlength{\tabcolsep}{3.2pt}
\renewcommand{\arraystretch}{1.08}

\caption{{Vessel segmentation results (mean $\pm$ std) on the internal and external datasets.
The best performance for each metric is highlighted in bold.
$***$ indicates that the difference in DSC between the proposed method and the state-of-the-art method is statistically significant at $p<0.001$ according to the Wilcoxon signed-rank test.
$\dagger$ indicates that manual annotations were used during training.}}

\resizebox{\textwidth}{!}{%
\begin{tabular}{l|ccccc|ccccc}
\toprule
\multirow{2}{*}{\textbf{Method}}
& \multicolumn{5}{c|}{\textbf{Internal}}
& \multicolumn{5}{c}{\textbf{External}} \\
\cmidrule(lr){2-6} \cmidrule(lr){7-11}

& \textbf{Acc.}
& \textbf{Sen.}
& \textbf{DSC}
& \textbf{clDice}
& \textbf{clRec.}
& \textbf{Acc.}
& \textbf{Sen.}
& \textbf{DSC}
& \textbf{clDice}
& \textbf{clRec.} \\
\midrule

nnUNet$^{\dagger}$
& 0.952 $\pm$ 0.048
& 0.580 $\pm$ 0.142
& 0.635 $\pm$ 0.128
& 0.684 $\pm$ 0.106
& 0.692 $\pm$ 0.035
& 0.957 $\pm$ 0.062
& 0.529 $\pm$ 0.152
& 0.620 $\pm$ 0.201
& 0.665 $\pm$ 0.168
& 0.671 $\pm$ 0.028 \\

MoG-RPCA
& 0.963 $\pm$ 0.058
& 0.774 $\pm$ 0.134
& 0.747 $\pm$ 0.140
& 0.782 $\pm$ 0.115
& 0.802 $\pm$ 0.028
& 0.943 $\pm$ 0.017
& 0.393 $\pm$ 0.152
& 0.494 $\pm$ 0.144
& 0.528 $\pm$ 0.121
& 0.548 $\pm$ 0.025 \\

MCR-RPCA
& 0.968 $\pm$ 0.013
& 0.476 $\pm$ 0.228
& 0.588 $\pm$ 0.197
& 0.612 $\pm$ 0.164
& 0.624 $\pm$ 0.042
& 0.918 $\pm$ 0.025
& 0.020 $\pm$ 0.032
& 0.031 $\pm$ 0.046
& 0.038 $\pm$ 0.042
& 0.041 $\pm$ 0.018 \\

TRPCA
& 0.965 $\pm$ 0.046
& 0.648 $\pm$ 0.189
& 0.695 $\pm$ 0.176
& 0.729 $\pm$ 0.143
& 0.744 $\pm$ 0.031
& 0.933 $\pm$ 0.024
& 0.164 $\pm$ 0.082
& 0.267 $\pm$ 0.119
& 0.291 $\pm$ 0.098
& 0.301 $\pm$ 0.022 \\

ETRPCA
& 0.955 $\pm$ 0.023
& 0.724 $\pm$ 0.097
& 0.654 $\pm$ 0.110
& 0.718 $\pm$ 0.088
& 0.767 $\pm$ 0.026
& 0.942 $\pm$ 0.017
& 0.681 $\pm$ 0.118
& 0.637 $\pm$ 0.069
& 0.695 $\pm$ 0.061
& 0.732 $\pm$ 0.019 \\

AngioNet
& 0.909 $\pm$ 0.056
& 0.500 $\pm$ 0.298
& 0.371 $\pm$ 0.202
& 0.415 $\pm$ 0.178
& 0.432 $\pm$ 0.045
& 0.880 $\pm$ 0.039
& 0.346 $\pm$ 0.186
& 0.297 $\pm$ 0.152
& 0.334 $\pm$ 0.131
& 0.361 $\pm$ 0.031 \\

Pix2pix
& 0.962 $\pm$ 0.028
& 0.718 $\pm$ 0.091
& 0.692 $\pm$ 0.106
& 0.715 $\pm$ 0.082
& 0.724 $\pm$ 0.030
& 0.945 $\pm$ 0.017
& 0.737 $\pm$ 0.096
& 0.670 $\pm$ 0.061
& 0.708 $\pm$ 0.049
& 0.746 $\pm$ 0.020 \\

DDSA
& 0.909 $\pm$ 0.061
& 0.679 $\pm$ 0.150
& 0.483 $\pm$ 0.127
& 0.552 $\pm$ 0.112
& 0.612 $\pm$ 0.038
& 0.893 $\pm$ 0.075
& 0.575 $\pm$ 0.191
& 0.477 $\pm$ 0.178
& 0.536 $\pm$ 0.151
& 0.586 $\pm$ 0.032 \\

C-DARL
& 0.962 $\pm$ 0.053
& 0.553 $\pm$ 0.113
& 0.658 $\pm$ 0.065
& 0.686 $\pm$ 0.058
& 0.703 $\pm$ 0.027
& 0.903 $\pm$ 0.074
& 0.510 $\pm$ 0.075
& 0.618 $\pm$ 0.057
& 0.643 $\pm$ 0.052
& 0.661 $\pm$ 0.024 \\

DOSTA-Net
& 0.977 $\pm$ 0.009
& 0.786 $\pm$ 0.088
& 0.789 $\pm$ 0.068
& 0.821 $\pm$ 0.054
& 0.842 $\pm$ 0.024
& 0.955 $\pm$ 0.011
& 0.761 $\pm$ 0.106
& 0.716 $\pm$ 0.065
& 0.754 $\pm$ 0.058
& 0.789 $\pm$ 0.019 \\

TDBM$^{***}$
& \textbf{0.986 $\pm$ 0.005}
& \textbf{0.803 $\pm$ 0.069}
& \textbf{0.867 $\pm$ 0.043}
& \textbf{0.871 $\pm$ 0.032}
& \textbf{0.872 $\pm$ 0.021}
& \textbf{0.961 $\pm$ 0.011}
& \textbf{0.810 $\pm$ 0.084}
& \textbf{0.755 $\pm$ 0.049}
& \textbf{0.801 $\pm$ 0.038}
& \textbf{0.831 $\pm$ 0.016} \\

\bottomrule
\end{tabular}%
}

\label{tab-vessel-seg}
\end{table*}

\begin{figure*}[h]
    \centering{
    \includegraphics[width=1\linewidth]{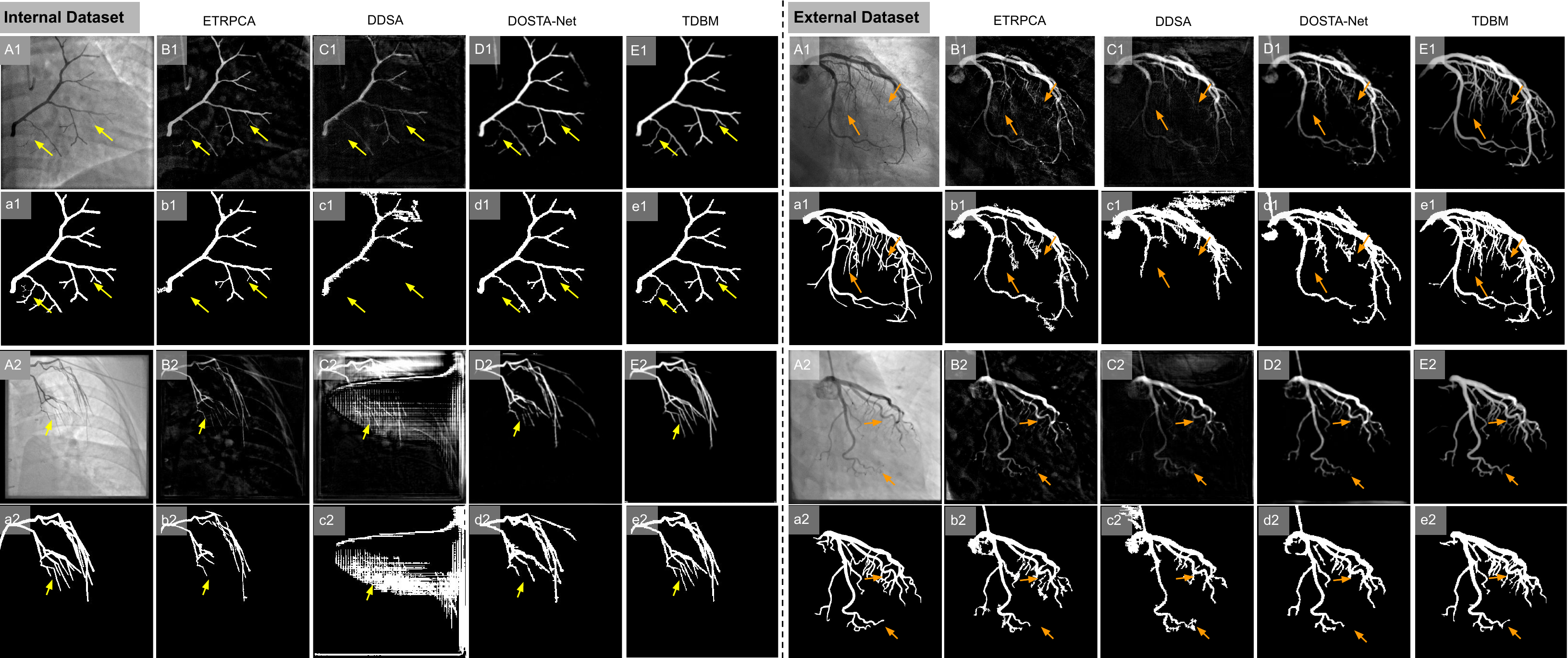}
    }
    \caption{The visualization results of vessel extraction and segmentation on the internal and external datasets are presented.  (A) and (a) show the input images and the corresponding ground truth of segmentation, respectively. (B–E) display the vessel extraction results obtained by ETRPCA, DDSA, DOSTA-Net and our TDBM. The corresponding binary segmentation results for each method are shown in (b–e).}
   \vspace{-5pt}
    \label{fig-ResVessel}
\end{figure*}

\subsection{Vessel Segmentation Comparison with Baseline Methods}
We rigorously benchmarked the proposed TDBM against a comprehensive set of state-of-the-art vessel extraction techniques, including training-free methods, supervised networks, and self-supervised approaches. The comparison set includes MoG-RPCA~\citep{zhao2014robust}, MCR-RPCA~\citep{jin2017extracting}, TRPCA~\citep{lu2019tensor}, ETRPCA~\citep{gao2020enhanced}, AngioNet~\citep{iyer2021angionet}, Pix2pix~\citep{isola2017image}, DDSA~\citep{duan2024training}, C-DARL~\citep{kim2024c}, and DOSTA-Net~\citep{hao2026dosta}.
The baselines represent diverse methodologies: MoG-RPCA, MCR-RPCA, TRPCA, and ETRPCA are training-free methods that exploit the low-rank structure of video sequences without requiring labeled data. In contrast, AngioNet, Pix2pix and DOSTA-Net are supervised methods trained directly on our synthetic data to ensure a fair comparison regarding data utilization. 
{AngioNet was included because it is a classical XCA vessel-segmentation method. However, AngioNet was originally trained using a private dataset and the authors did not release trained model weights. We therefore trained AngioNet using the same synthetic data as TDBM precisely to control for training data and isolate architecture/modeling differences. Using the same synthetic training data therefore provides a more controlled comparison.}
DDSA represents a domain adaptation approach, while C-DARL employs self-supervised contrastive learning. Furthermore, we established a fully-supervised performance baseline by training nnU-Net~\citep{isensee2021nnu} using the manually annotated DCA1 dataset~\citep{cervantes2019automatic}, recognizing its reliance on manual labels. The quantitative segmentation results, detailed in Table~\ref{tab-vessel-seg}, demonstrate the superior performance and robustness of the TDBM across two distinct evaluation scenarios.

On the internal dataset, the proposed TDBM significantly outperformed all compared baselines across all key metrics. Notably, TDBM achieved the best overall performance with a DSC of ${0.867} \pm {0.043}$, an Accuracy of ${0.986} \pm {0.005}$. This strong performance confirms the effectiveness of our physics-based data engine combined with the TDBM architecture in leveraging the temporal coherence and fine-grained anatomical realism of the synthetic sequences.

The model's generalization capability was validated on the challenging External Dataset (XCA30), which contains real clinical artifacts and noise characteristics. Here, TDBM maintained its dominance, achieving a DSC of ${0.755} \pm {0.049}$, alongside the highest Accuracy (${0.961} \pm {0.011}$) and Sensitivity (${0.810} \pm {0.084}$) among all compared methods. Critically, the performance improvement of TDBM over the next-best method, DOSTA-Net (DSC $0.716$), was confirmed to be statistically significant ($p < 0.001$) via the Wilcoxon signed-rank test.

{To further assess generalizability on a widely used public XCA benchmark, we additionally evaluated TDBM on all 134 images of the DCA1 dataset~\citep{cervantes2019automatic}. As shown in Table~\ref{tab:dca1_response}, we compared TDBM with several recent methods reported on the same dataset, including ~\citep{ferrari2025corosam}, FAVS~\citep{yuan13hospital}, TRACS~\citep{kaushal2026tracs}, and DOSTA-Net~\citep{hao2026dosta}. To make the different supervision settings explicit, we report whether each method requires manual vessel annotations and the number of labeled real XCA images used for training or fine-tuning. TDBM achieved a Dice score of 0.671 without using any manually annotated real XCA images, outperforming the annotation-free TRACS (0.616) and DOSTA-Net (0.282), while remaining competitive with the supervised CoroSAM (0.730) and FAVS (0.705), which use 1,200 and 1,326 labeled training images, respectively. Results for CoroSAM, FAVS, and TRACS are taken directly from their original publications to avoid potential bias introduced by reimplementation or retraining.}

\begin{table}[H]
\scriptsize
\centering
{
\caption{{Comparison of vessel segmentation performance on the DCA1 dataset. ``Labeled training data'' denotes the number of real XCA images with manual vessel annotations used for training or fine-tuning.}}
\label{tab:dca1_response}
\begin{tabular}{lccccc}
\toprule
Method & Year & Training paradigm & Labeled training data & Dice/F1 \\
\midrule
CoroSAM$^{a}$ & 2026 & Supervised & 1,200 & 0.730 \\
FAVS$^{a}$ & 2026 & Supervised & 1,326 & 0.705 \\
TRACS$^{a}$ & 2026 & Annotation-free & 0 & 0.616 \\
DOSTA-Net$^{b}$ & 2026 & Annotation-free & 0 & 0.282 \\
TDBM (Ours) & 2026 & Annotation-free & 0 & 0.671 \\
\bottomrule
\end{tabular}
\vspace{1mm}
\begin{minipage}{0.96\linewidth}
\footnotesize $^{a}$Results are quoted directly from the corresponding original publications; these methods were not retrained or re-evaluated by us.\\
$^{b}$The DOSTA-Net result was obtained by direct inference using the authors' released multi-frame checkpoint, without retraining, fine-tuning, or target-domain adaptation. Because DCA1 contains only single-frame angiograms, each image was replicated three times to construct the required temporal input. The lack of genuine temporal variation likely contributes to the relatively low performance in this setting.
\end{minipage}
}
\end{table}

{Taken together, the results on XCA30 and DCA1 demonstrate that TDBM maintains strong cross-dataset generalization without requiring manually annotated real XCA images for training, while achieving performance competitive with recent supervised approaches.}

The representative visualization results illustrating the efficacy of vessel extraction and segmentation by the proposed methods and baselines are presented in $\text{Fig. }\ref{fig-ResVessel}$.
A qualitative inspection reveals that our TDBM consistently outperforms existing approaches. Specifically, we observe that TDBM is able to detect subtler and finer vessel structures, which are often missed or poorly segmented by competitor methods. Furthermore, TDBM achieves this high recall while simultaneously exhibiting significantly less background noise and fewer false positives in the segmented masks compared to other techniques, resulting in clearer and more anatomically accurate vessel representations.

\subsection{Reader Study}
\begin{figure}[ht]
    \centering{
    \includegraphics[width=1\linewidth]{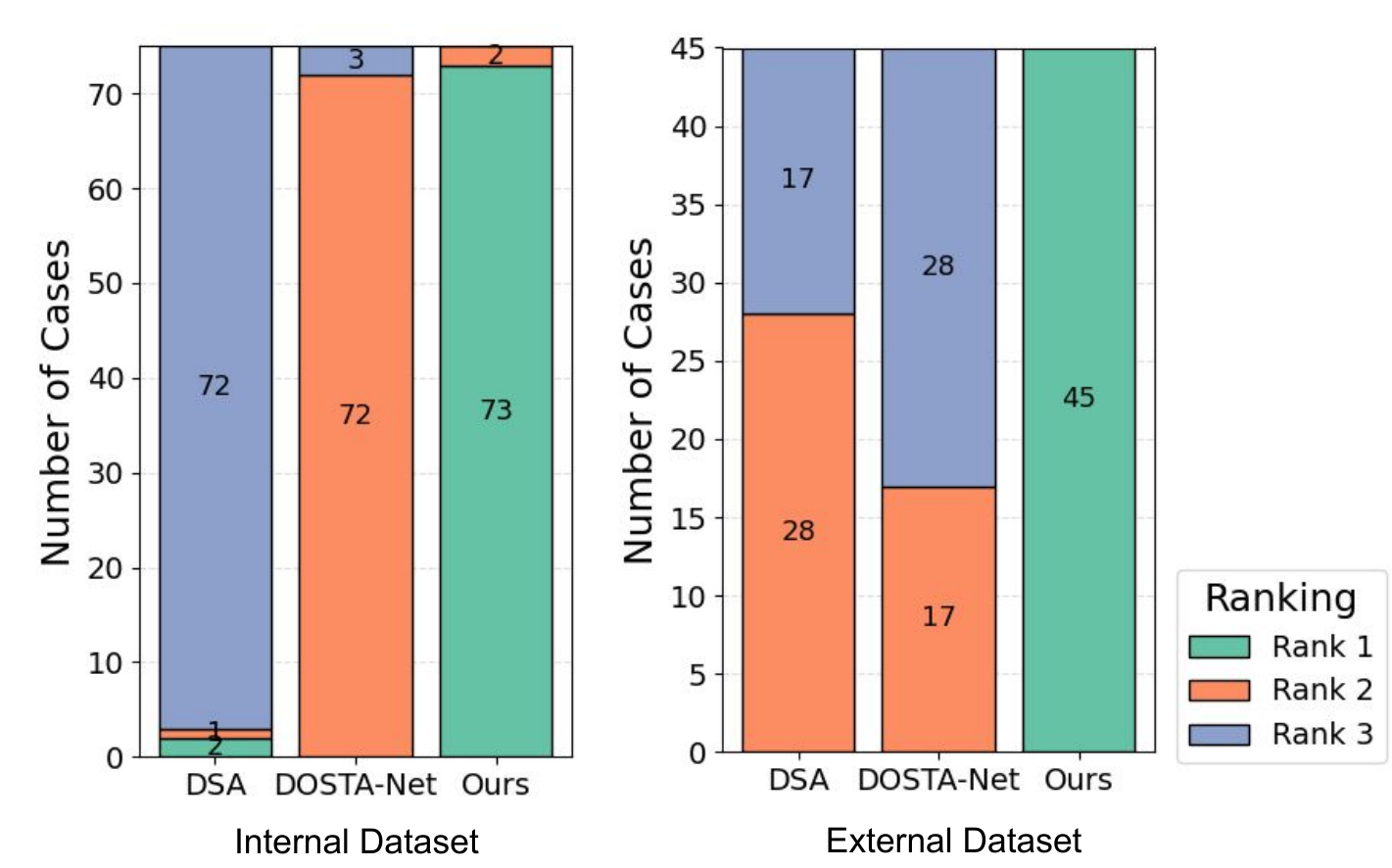}
    }
\caption{{Overall ranking distribution of the reader study across internal and external datasets. Three board certified radiologists independently ranked the results of each method (DSA, DOSTA-Net, and Ours) based on vessel extraction quality.}}
    \label{fig-ranking}
\end{figure}

To evaluate the clinical relevance of the proposed method, we conducted a reader study on both the internal and external datasets. Vessel extraction was performed using three different methods: conventional Digital Subtraction Angiography (DSA), the state-of-the-art deep learning baseline DOSTA-Net, and our proposed TDBM. For each case, the results from the three methods were randomized and labeled as R1, R2, and R3 to blind the evaluators to the source method. Three board-certified radiologists (authors D. R. Cantrell, R. Abdalla, and S. A. Ansari) independently reviewed the images. They were asked to provide an overall ranking (Rank 1: Best, Rank 2: Intermediate, Rank 3: Worst) for the vessel extraction quality of the three methods for each case.
The distribution of the rankings is summarized in Fig.~\ref{fig-ranking}. The results demonstrate an overwhelming preference for our proposed TDBM across both datasets:

\textbf{Internal Dataset:} Our method achieved the highest quality (Rank 1) in 73 out of 75 evaluations ($97.3\%$). DOSTA-Net generally performed as the second-best method (Rank 2 in 72 cases), while conventional DSA was consistently ranked lowest (Rank 3 in 72 cases), primarily due to motion artifacts and background noise.

\begin{figure*}[ht]
    \centering{
    \includegraphics[width=1\linewidth]{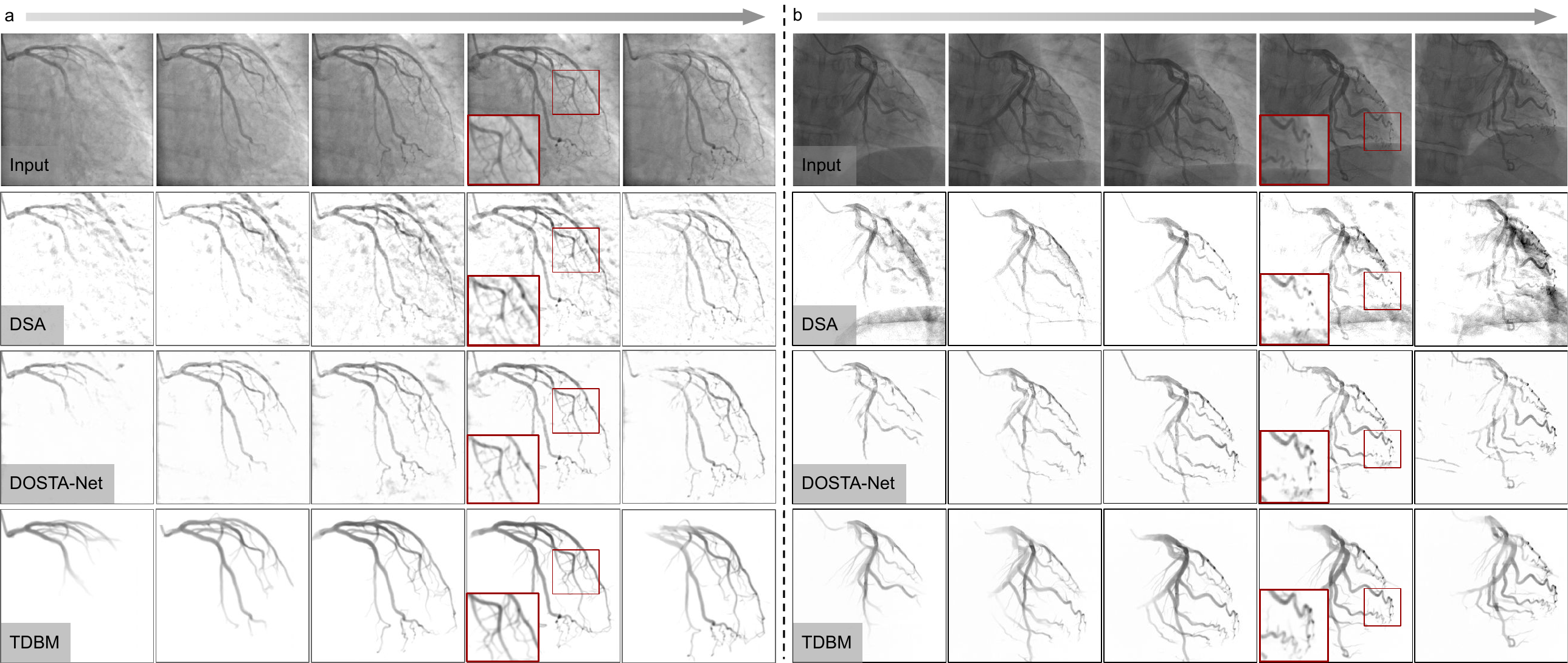}
    }
    \caption{Visualization results of real-sequence XCA images used for subjective evaluation from the internal (a) and external (b) datasets. Results obtained using DSA, DOSTA-Net, and the proposed TDBM are shown separately along the sequence. Red boxes highlight and provide magnified views of the key regions of interest.}
    \label{fig-Resreader}
\end{figure*}

\textbf{External Dataset:} The robustness of our method was further validated on the external dataset, where it was ranked as the best method in all 45 cases ($100\%$). Notably, although DOSTA-Net outperformed DSA on the internal dataset, its performance degraded substantially on the external dataset, ranking last (Rank 3) in 28 cases. In contrast, the conventional DSA method ranked second in the majority of external cases (28 cases), outperforming DOSTA-Net. This discrepancy may be attributed to reduced motion in the external dataset compared with the internal dataset, under which DSA is less affected and thus exhibits better performance than DOSTA-Net.
We also present the qualitative visualization results from the reader study dataset in Fig.~\ref{fig-Resreader}.


\subsection{Ablations and Other Analysis}
We conducted a series of ablation and analysis experiments to understand the effect of different design choices and training strategies in our framework. These experiments evaluate: the temporal augmentation strategy, the number of temporal slices used as model input, the Vessel-aware Contrastive Learning (VCL) module,  the annealing loss term for incorporating real data, and the impact of training with different amounts of synthetic data. The quantitative results on the external dataset for vessel segmentation are summarized in Table~\ref{tab-ablation}.

\textbf{{Effectiveness of Temporal Augmentation.}}
We first evaluate the impact of our proposed Temporal Augmentation. By comparing Model 2 (without augmentation) and Model 3 (with augmentation), we observe a clear performance gain. Applying temporal augmentation boosts the mean DSC from 0.738 to 0.755 and Sensitivity from 0.792 to 0.810. This demonstrates that our augmentation strategy effectively enriches the temporal context and improves the model's ability to capture vessel dynamics, leading to more accurate segmentation.

\textbf{{Influence of Input Slice Number.}}
To find the optimal temporal window, we experimented with varying numbers of input slices. First, comparing Model 1 (1 slice) and Model 2 (3 slices), both without augmentation, the DSC improves from 0.727 to 0.738. This indicates that a single frame provides insufficient temporal information, and incorporating even a small number of adjacent slices is beneficial.
We then analyzed the impact of slice numbers 3, 5, and 7, all with temporal augmentation (Models 3, 4, and 5). Model 3 (3 slices) achieves the highest DSC (0.755). Increasing the slice number to 5 (Model 4) and 7 (Model 5) resulted in a performance drop to 0.747 and 0.746, respectively. This suggests that excessive temporal context introduces noise or motion inconsistency, whereas three slices provide a balanced and informative temporal window.

\textbf{{Effectiveness of VCL and Annealing Term.}}
We validated the contribution of the VCL and the annealing loss term by removing them individually from our best model (Model 3). Removing the VCL component (Model 7) causes a significant performance drop, with the DSC decreasing from 0.755 to 0.736. This highlights the crucial role of VCL in learning discriminative vessel features. Similarly, removing the annealing term (Model 6) reduces the DSC to 0.743, confirming its utility in effectively integrating imperfect real data during the training process and stabilizing the diffusion bridge.

\textbf{{Influence of Training Data Volume.}}
Rows 3, 8, and 9 compare models trained with 50k, 20k, and 5k synthetic samples, respectively. Performance consistently improves with more synthetic data, with DSC increasing from 0.703 (5k samples) to 0.728 (20k samples) and peaking at 0.755 with 50k samples. These findings demonstrate that large-scale synthetic data substantially enhance generalization to unseen clinical domains, especially given the limited availability of high-quality real annotations.

\textbf{Influence of VCL location.}
We additionally evaluated VCL at the deep encoder level, bottleneck level, decoder path, and all selected levels jointly (Table~\ref{tab-vcl-location}). The layer-wise ablation showed that applying VCL at any individual feature level consistently improved performance over the no-VCL baseline. The deepest encoder block and bottleneck increased DSC from 0.736 to 0.743 and 0.746, respectively, indicating that high-level contextual representations are beneficial for vessel discrimination. Applying VCL to the decoder stages yielded a larger improvement (DSC = 0.751), suggesting that higher-resolution decoder features are particularly useful for preserving fine vessel structures. The best performance was obtained when VCL was jointly applied across all selected encoder, bottleneck, and decoder levels (DSC = 0.755), supporting the complementary role of multi-scale contextual and spatially detailed representations.

\begin{table*}[h]
\scriptsize
\centering
\setlength{\tabcolsep}{3.5pt}
\renewcommand{\arraystretch}{1.08}

\caption{{Performance (mean $\pm$ std) of the ablation study on the external vessel segmentation dataset. 
$\star$ indicates the default configuration used throughout the main experiments.}}

\resizebox{\textwidth}{!}{%
\begin{tabular}{c|ccccc|ccccc}
\hline
No. 
& \makecell[c]{Temporal\\Cond.}
& \makecell[c]{Slice\\No.}
& VCL
& \makecell[c]{Annealing\\Term}
& \makecell[c]{Training\\No.}
& Acc.
& Sen.
& DSC
& clDice
& clRec. \\
\hline

1 
& \ding{55} 
& 1 
& \ding{51} 
& \ding{51} 
& 50k
& 0.958 $\pm$ 0.012 
& 0.776 $\pm$ 0.151 
& 0.727 $\pm$ 0.094
& 0.744 $\pm$ 0.091
& 0.801 $\pm$ 0.126 \\

2 
& \ding{55} 
& 3 
& \ding{51} 
& \ding{51} 
& 50k
& 0.957 $\pm$ 0.014 
& 0.792 $\pm$ 0.096 
& 0.738 $\pm$ 0.060
& 0.758 $\pm$ 0.061
& 0.814 $\pm$ 0.091 \\

\textbf{3$^\star$} 
& \ding{51} 
& 3 
& \ding{51} 
& \ding{51} 
& 50k
& \textbf{0.961 $\pm$ 0.011}
& \textbf{0.810 $\pm$ 0.084}
& \textbf{0.755 $\pm$ 0.049}
& \textbf{0.779 $\pm$ 0.051}
& \textbf{0.831 $\pm$ 0.078} \\

4 
& \ding{51} 
& 5 
& \ding{51} 
& \ding{51} 
& 50k
& 0.954 $\pm$ 0.016 
& 0.795 $\pm$ 0.058 
& 0.747 $\pm$ 0.056
& 0.768 $\pm$ 0.057
& 0.819 $\pm$ 0.060 \\

5 
& \ding{51} 
& 7 
& \ding{51} 
& \ding{51} 
& 50k
& 0.949 $\pm$ 0.014 
& 0.798 $\pm$ 0.050 
& 0.746 $\pm$ 0.048
& 0.766 $\pm$ 0.050
& 0.821 $\pm$ 0.055 \\

6 
& \ding{51} 
& 3 
& \ding{51} 
& \ding{55} 
& 50k
& 0.953 $\pm$ 0.015 
& 0.808 $\pm$ 0.052 
& 0.743 $\pm$ 0.052
& 0.762 $\pm$ 0.054
& 0.823 $\pm$ 0.057 \\

7 
& \ding{51} 
& 3 
& \ding{55} 
& \ding{51} 
& 50k
& 0.954 $\pm$ 0.014 
& 0.790 $\pm$ 0.050 
& 0.736 $\pm$ 0.048
& 0.748 $\pm$ 0.051
& 0.803 $\pm$ 0.057 \\

8 
& \ding{51} 
& 3 
& \ding{51} 
& \ding{51} 
& 5k
& 0.954 $\pm$ 0.012 
& 0.739 $\pm$ 0.147 
& 0.703 $\pm$ 0.085
& 0.719 $\pm$ 0.084
& 0.767 $\pm$ 0.125 \\

9 
& \ding{51} 
& 3 
& \ding{51} 
& \ding{51} 
& 20k
& 0.955 $\pm$ 0.014 
& 0.793 $\pm$ 0.118 
& 0.728 $\pm$ 0.062
& 0.747 $\pm$ 0.064
& 0.809 $\pm$ 0.101 \\

\hline
\end{tabular}%
}

\label{tab-ablation}
\end{table*}

\begin{table}[h] 
\centering
\caption{{Ablation on the location of the vessel-aware contrastive loss (VCL) on the external XCA30 dataset (mean $\pm$ std). }}
\label{tab-vcl-location}
\resizebox{\columnwidth}{!}{%
\begin{tabular}{l|l|ccccc}
\hline
Setting & VCL location   & DSC & clDice \\ \hline

No VCL        
& None                                

& 0.736 $\pm$ 0.048 
& 0.748 $\pm$ 0.051 \\

Deep encoder  
& Deepest encoder block (1/8)         

& 0.743 $\pm$ 0.050 
& 0.758 $\pm$ 0.052 \\

Bottleneck    
& Bottleneck (1/8)                    

& 0.746 $\pm$ 0.051 
& 0.762 $\pm$ 0.053 \\

Decoder       
& Three decoder stages                
 
& 0.751 $\pm$ 0.050 
& 0.771 $\pm$ 0.052 \\

All levels    
& Encoder + bottleneck + decoder      

& 0.755 $\pm$ 0.049 
& 0.779 $\pm$ 0.051 \\ \hline
\end{tabular}}
\end{table}

\begin{figure}[ht]
    \centering{
    \includegraphics[width=1\linewidth]{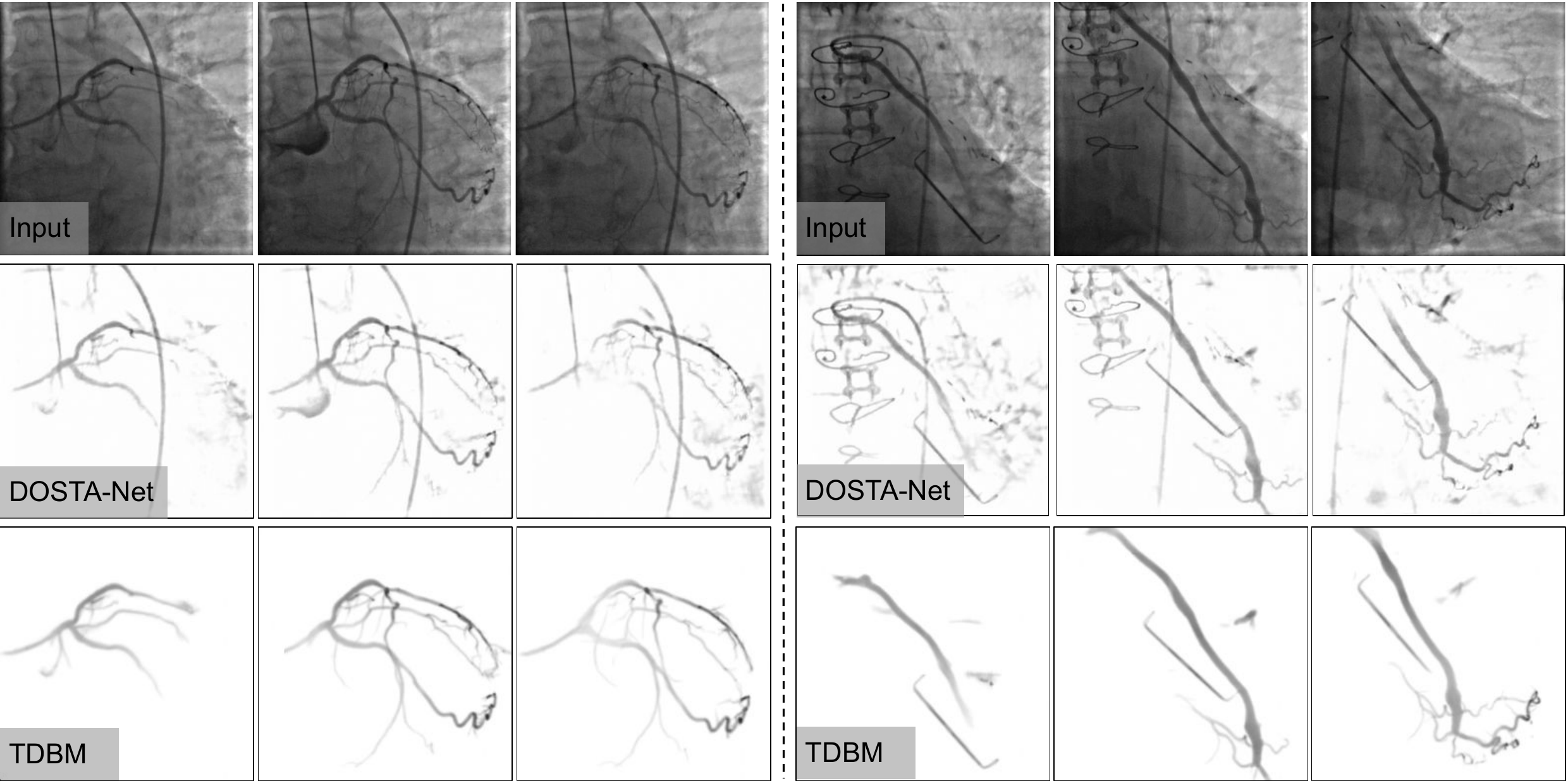}
    }
\caption{{Visualizations of representative challenging cases involving interventional artifacts.}}
    \label{fig-failed}
\end{figure}

\subsection{Sampling Acceleration Analysis}

As a diffusion model, the computational inefficiency of the standard $N=1000$ sequential noise removal steps limits the practical utility of TDBM. To enhance inference speed, we applied the accelerated sampling technique of Denoising Diffusion Implicit Models~\citep{songdenoising}, which utilizes a reduced subset of timesteps ($K$).

We systematically tested accelerated sampling rates up to $500\times$ (relative to $N=1000$ steps) on the external XCA dataset, as shown in Table~\ref{tab-acceleration}. Our analysis confirms that TDBM maintains near-maximum performance even under highly sparse sampling schedules. Specifically, reducing the steps from $N=1000$ to $K=50$ (a $20\times$ acceleration) resulted in a negligible change in the Dice Similarity Coefficient (DSC $0.755 \pm 0.050$ vs. $0.755 \pm 0.049$). Furthermore, a $100\times$ acceleration using $K=10$ steps still yielded a robust DSC of $0.753 \pm 0.053$. Performance only degraded significantly at the extreme $500\times$ acceleration ($K=2$ steps, DSC $0.159 \pm 0.106$). In conclusion, TDBM can achieve its optimal segmentation quality with a $\mathbf{20\times}$ reduction in inference time by using $K=50$ DDIM steps.

{In addition, we conducted a computational-efficiency comparison reporting parameter count, sampling steps, GPU memory, latency, and DSC. As shown in Table~\ref{tab:efficiency}, the iterative nature of TDBM makes it slower than single-pass models such as pix2pix and nnU-Net. However, accelerated sampling substantially reduces this cost while preserving segmentation performance. Specifically, reducing the sampling steps from 1000 to 50 decreases the latency from 20.09 to 1.00 s/frame with essentially unchanged DSC (0.755 vs.\ 0.755), while 10 steps further reduces the latency to approximately 0.20 s/frame with only a minor decrease in DSC to 0.753.}

\begin{table}[htbp]
\renewcommand\arraystretch{1.2}
\centering
\caption{{Effect of accelerated sampling steps on vessel segmentation performance over the external XCA dataset.}}
\label{tab-acceleration}
\resizebox{\columnwidth}{!}{%
\begin{tabular}{ccccc}
\hline
\textbf{Sampling Steps} & \textbf{Accuracy} & \textbf{Sensitivity } & \textbf{Specificity} & \textbf{DSC} \\ 
\hline
2 & 0.928 $\pm$ 0.024 & 0.093 $\pm$ 0.069 & 0.999 $\pm$ 0.002 & 0.159 $\pm$ 0.106 \\
10 & 0.961 $\pm$ 0.011 & 0.799 $\pm$ 0.096 & 0.973 $\pm$ 0.013 & 0.753 $\pm$ 0.053  \\
50 & 0.961 $\pm$ 0.010 & 0.806 $\pm$ 0.095 & 0.973 $\pm$ 0.013 & 0.755 $\pm$ 0.050  \\
200 & 0.960 $\pm$ 0.011 & 0.807 $\pm$ 0.088 & 0.972 $\pm$ 0.014 & 0.753 $\pm$ 0.049  \\
1000 & {0.961} $\pm$ 0.011 & {0.810} $\pm$ 0.084 & {0.972} $\pm$ 0.013 &{0.755} $\pm$ 0.049 \\
\hline
\end{tabular}
}
\end{table}

\begin{table}[!h]
\small
\centering
\caption{Computational efficiency and segmentation performance on the external XCA dataset.}
\resizebox{\columnwidth}{!}{
\begin{tabular}{lccccc}
\hline
Method & Params (M) & Steps & VRAM (GB) & Latency (s/frame) & DSC \\
\hline
TDBM & 62.6 & 1000 & 0.77 & 20.09 & 0.755 $\pm$ 0.049 \\
TDBM & 62.6 & 50 & 0.77 & 1.00 & 0.755 $\pm$ 0.050 \\
TDBM & 62.6 & 10 & 0.77 & 0.20 & 0.753 $\pm$ 0.053 \\
pix2pix & 54.4 & - &  0.45 & 0.04 & 0.670 $\pm$ 0.061 \\
nnU-Net & 20.6 & - & 0.21 & 0.04 & 0.620 $\pm$ 0.201 \\
\hline
\end{tabular}
}
\label{tab:efficiency}
\end{table}

\subsection{Zero-shot segmentation on other modalities}
To assess the cross-modality generalization of TDBM, we evaluate its zero-shot vessel segmentation performance on two unseen modalities: DRIVE and VessMAP. We compare TDBM with MedSegX~\citep{zhang2025generalist}, a recently proposed medical foundation model, and a fully supervised nnUNet trained on each target dataset.
{Because TDBM was trained at $256\times256$ due to computational constraints, all methods, including nnU-Net, were trained and evaluated using the same $256\times256$ images to ensure a fair within-study comparison.}

As shown in Table 8, TDBM achieves strong zero-shot performance on DRIVE, performing on par with, and slightly above, the fully supervised nnUNet trained on the target dataset (DSC: 0.739 ± 0.044 vs. 0.731 ± 0.013; Sensitivity: 0.789 vs. 0.739), and outperforming MedSegX by a large margin across all metrics. On VessMAP, TDBM substantially surpasses the zero-shot foundation model MedSegX (DSC: 0.773 vs. 0.591), although a clear gap remains relative to the supervised nnUNet (DSC: 0.919), which is trained on labeled data from the target domain. These results indicate that TDBM transfers effectively to unseen modalities without any target-domain annotations, while supervised in-domain training remains advantageous when labeled data are available.

\begin{figure}[ht]
    \centering{
    \includegraphics[width=1\linewidth]{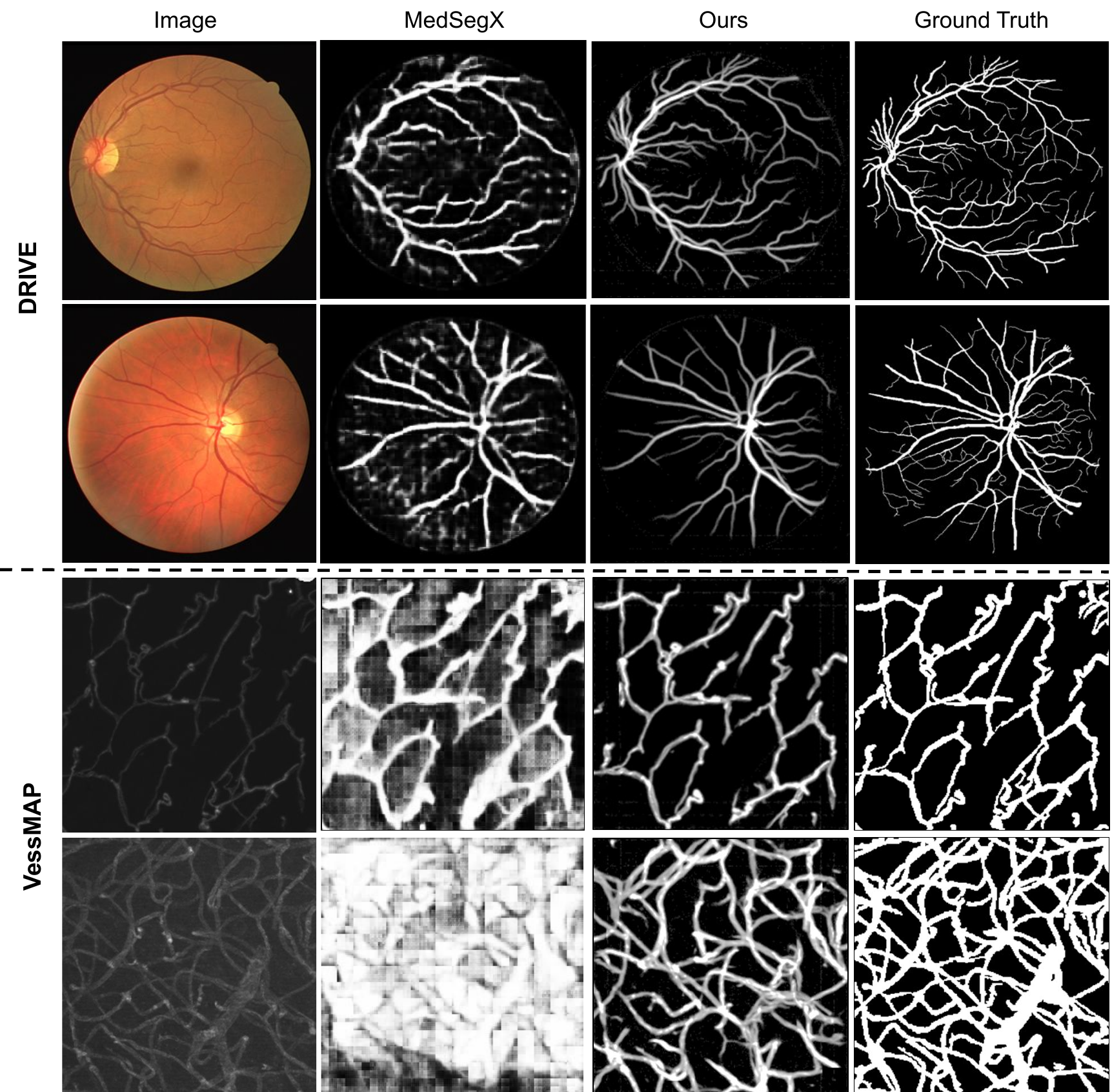}
    }
\caption{{Visualizations of zero-shot vessel segmentation performance on DRIVE and VessMAP datasets.}}
    \label{fig-zero-shot}
\end{figure}

Figure~\ref{fig-zero-shot} further illustrates representative results. TDBM produces coherent and complete vessel structures across both modalities. Minor omissions of fine vessels on DRIVE can be attributed to the required downsampling to 256$\times$256 during inference, which may suppress small-caliber vessels.

\begin{table}[htbp]
\centering
\caption{Zero-shot Segmentation Performance Comparison on DRIVE and VessMAP datasets. $\dagger$ indicates baselines trained in a supervised manner using labeled data.}
\label{tab-zero-shot}
\resizebox{\columnwidth}{!}{%
\begin{tabular}{llccc}
\hline
\textbf{Dataset} & \textbf{Method} & \textbf{Accuracy} & \textbf{Sensitivity} & \textbf{DSC} \\
\hline
\multirow{3}{*}{DRIVE} 
& nnUNet$^{\dagger}$ & 0.953 $\pm$ 0.003 & 0.739 $\pm$ 0.044 & 0.731 $\pm$ 0.013 \\
& MedSegX & 0.918 $\pm$ 0.005 & 0.581 $\pm$ 0.043 & 0.551 $\pm$ 0.030 \\
& Ours & {0.955 $\pm$ 0.006} & {0.789 $\pm$ 0.086} & {0.739 $\pm$ 0.044} \\
\hline
\multirow{3}{*}{VessMAP} 
& nnUNet$^{\dagger}$ & {0.960 $\pm$ 0.020} & {0.942 $\pm$ 0.040} & {0.919 $\pm$ 0.037} \\
& MedSegX & 0.644 $\pm$ 0.272 & 0.782 $\pm$ 0.130 & 0.591 $\pm$ 0.184 \\
& Ours & 0.867 $\pm$ 0.076 & 0.763 $\pm$ 0.133 & 0.773 $\pm$ 0.075 \\
\hline
\end{tabular}%
}
\end{table}

\section{Discussion and Conclusion}
In this study, we presented the Temporal Diffusion Bridge Model (TDBM), a framework that reframes coronary artery extraction from a traditional segmentation task into an image translation problem. By leveraging a mechanistic synthesis engine and a temporally-conditioned diffusion bridge, our approach overcomes the long-standing reliance on labor-intensive human annotations and the inherent failures of conventional DSA in the presence of cardiac and respiratory motion.

Unlike purely data-driven generative models (e.g., GANs) that often produce hallucinated vascular structures, our engine utilizes Murray’s Law and Gamma-variate flow dynamics to ensure the synthesized vessels are coherent. This adherence to physical principles likely explains the model’s remarkable zero-shot generalization to retinal (DRIVE) and microscopy (VessMAP) datasets. Even though these modalities differ in texture and scale, the underlying tubular geometry and contrast-to-background relationships are consistent with the physical priors encoded in our training data. This suggests that physics-based synthesis can effectively bridge the domain gap by teaching the model the fundamental essence of vascular structures rather than just memorizing specific X-ray artifacts.

{The choice of a Diffusion Bridge over standard denoising diffusion probabilistic models (DDPM) or U-Nets is pivotal. By formulating the process as a Brownian Bridge between contrast-filled and contrast-free states, the model learns a directed mapping that preserves the fine-grained intra-arterial contrast dynamics. This is a significant departure from binary segmentation, as it retains the pixel-level intensity variations that are vital for subsequent hemodynamic analysis and stenosis quantification.} Furthermore, our temporal conditioning strategy effectively addresses the "jitter" often seen in frame-by-frame deep learning models. By observing motion cues, the TDBM learns to distinguish between moving contrast agents and static or quasi-periodic background anatomy, achieving a level of temporal consistency.

Despite the robust performance, certain challenges remain. As shown in our analysis of challenging cases (Fig. \ref{fig-failed}), the presence of dense interventional hardware, such as large-bore catheters or metallic stents, can occasionally lead to extraction artifacts. Since our current synthesis engine does not explicitly model the complex X-ray physics of metallic implants, the model may misinterpret these as vascular structures or background noise. 
{Randomly projecting synthetic coronary trees onto real backgrounds can produce anatomically imperfect combinations and contributes to the domain gap between synthetic and real XCA.
However, our objective is not to generate synthetic images that are anatomically identical to real XCA on a case-by-case basis. Instead, the synthetic pipeline is designed to generate sufficiently diverse vessel appearances and background combinations so that the model learns to distinguish transient vessel-related contrast from persistent background structures. The remaining synthetic-to-real gap is further reduced using unlabeled real XCA sequences during training.}
Future iterations of the data engine will incorporate 3D models of interventional devices to improve robustness in complex procedural settings. 

In conclusion, the proposed TDBM framework demonstrates that high-fidelity vessel extraction can be achieved without the need for real-world manual labels. By combining the rigor of physics-based modeling with the generative power of diffusion bridges, we provide a robust solution for XCA analysis that generalizes across datasets and modalities. The strong preference shown by clinical experts and the competitive quantitative results against supervised baselines underscore the potential of this framework to be integrated into the next generation of automated cardiovascular diagnostic systems.

\bibliographystyle{model2-names.bst}\biboptions{authoryear}
\bibliography{refs}
\end{document}